\documentclass[a4paper,11pt]{JHEP3}
\usepackage{epsf}
\newcommand{\be}{\begin{eqnarray}}
\newcommand{\ee}{\end{eqnarray}}
\newcommand{\abs}[1]{\left| #1 \right|}
\newcommand{\bra}[1]{\langle{ #1}|}
\newcommand{\ket}[1]{|{ #1}\rangle}
\newcommand{\ip}[2]{\langle{ #1}|{ #2}\rangle}
\newcommand{\op}[2]{\left|\left.{ #1\,}\right\rangle\right.\!\!
                    \left\langle\left.{ #2\,}\right|\right.}
\newcommand{\tr}{{\rm Tr}}
\newcommand{\str}{{\rm tr}}
\newcommand{\dg}{\dagger}
\renewcommand{\Re}{{\mathcal Re}}
\renewcommand{\Im}{{\mathcal Im}}

\author{
Leszek Hadasz\footnote{
M. Smoluchowski Institute of Physics,
    Jagiellonian University,
    Reymonta 4, 30-059 Krak\'ow, Poland}
\\
    Laboratoire de Physique Th\`{e}orique, \\
    Univ. Paris-Sud, B{\^a}t. 210, F-91405 Orsay Cedex \\
\email{hadasz@th.u-psud.fr}
}
\author{
Ulf Lindstr\"{o}m \\
Department of Theoretical Physics,\\
Uppsala University, Box 803, SE-751 08 Uppsala, Sweden\\
\email{ulf.lindstrom@teorfys.uu.se}
}
\author{
Martin Ro\v{c}ek \\
State University of New York\\
Stony Brook, NY 11794-3840, USA \\
\email{rocek@insti.physics.sunysb.edu}
}
\author{
Rikard von Unge \\
Institute for Theoretical Physics and Astrophysics\\
Faculty of Science, Masaryk University\\
Kotl\'{a}\v{r}sk\'{a} 2, CZ-611 37, Brno, Czech Republic\\
\email{unge@physics.muni.cz}
}

\abstract{We study one- and two-soliton solutions of noncommutative
Chern-Simons theory coupled to a nonrelativistic or a relativistic
scalar field. In the nonrelativistic case, we find a tower of new stationary time-dependent solutions, all with the same charge density, but with increasing energies. The dynamics of these solitons cannot be studied using traditional moduli space techniques, but we do find a nontrivial symplectic form on the phase space indicating that the moduli space is not flat. In the relativistic case we find the metric on the two soliton moduli space.}

\title{Time dependent solitons of noncommutative Chern-Simons theory
coupled to scalar fields}
\preprint{UUITP-13-03\\
YITP-SB-03-38\\
hep-th/0309015}

\begin{document}

\section{Introduction}
Soliton solutions of noncommutative field theory (for
an introduction to the subject and references
see for instance
\cite{Seiberg:1999vs, Harvey:2001yn, Douglas:2001ba, Szabo:2001kg})
is a fascinating
subject. In some cases, theories that do not have soliton solutions
classically turn out to have them when made noncommutative
\cite{Gopakumar:2000zd}. The fact
that these solitons have an interpretation as lower dimensional
D-branes in the context of string-theory
\cite{Schomerus:1999ug, Seiberg:1999vs}
and as quasi-particles in the
context of noncommutative descriptions of solid state systems
\cite{Susskind:2001fb, Polychronakos:2001mi} makes
this phenomenon even more interesting. We therefore believe that it is
worthwhile to study solitonic configurations and their detailed
dynamics for a range of noncommutative theories \cite{LRvU,HLRvU}.

Noncommutative Chern-Simons theory is particularly interesting, as
Susskind has argued that it describes the fractional quantum Hall
effect \cite{Susskind:2001fb}, (see also \cite{Polychronakos:2001mi,
Hellerman:2001yv, Freivogel:2001vc,HK1,HK2}). In Susskind's description the
solitons are the quasiparticles that exhibit fractional statistics and
are responsible for the unusual quantum behavior of the QHE
system. According to Susskind, it is the noncommutativity of the
theory that encodes the graininess of the two-dimensional charged fluid in a magnetic field.

In this work we study soliton solutions of noncommutative
Chern-Simons theory coupled to relativistic and nonrelativistic
scalar fields \cite{Bak:2001sg, Lozano:2000md}; these theories are
noncommutative versions of the Jackiw-Pi model \cite{Jackiw:1990mb,
Jackiw:1990md}.  We find the shape of two-soliton configuration using
the exact Seiberg-Witten map \cite{sw1,sw2,sw3,sw4}. We also find new stationary time-dependent single soliton solutions, all with the same $\delta$-function charge density profile, but with increasing energies. Finally, in
the relativistic case, we are able to find the dynamics of the two
soliton solution using traditional moduli space techniques.

We think that our results are novel and striking; however, in contrast to the case of pure noncommutative Chern-Simons theory, the details of the physical interpretation, in particular the relation to the fractional quantum Hall effect, remain to be sorted out.

This paper is a thoroughly revised version. We thank P.~A.~Horvathy, L.~Martina and P.~C.~Stichel for illuminating comments that led us to reconsider some of our previously reported results; see note added at the end of the paper.

\section{Pure Chern-Simons}
We take our noncommutative space to be defined by the commutation
relation $\left[\hat{x},\hat{y}\right] = -i\theta$. Noncommutative
Chern-Simons theory is defined by the Lagrangian
\be
 L = -\pi\kappa\theta \epsilon^{\alpha\beta\gamma} \tr\left(
 A_\alpha\partial_\beta A_\gamma  -\frac{2i}{3}A_\alpha A_\beta
A_\gamma \right)~,
\ee
where $\kappa$ is the Chern-Simons level, representing the (inverse) filling fraction in the application to the fractional quantum Hall effect\footnote{In what follows we shall need to take $\kappa$ positive to find our solutions.}.
Introducing the notation $X^{i} = x^{i} - \theta\epsilon_{ij}A_j$ we
write the Lagrangian as \cite{Brodie:2000yz,Polychronakos:2000nt,Kluson:2000mv}
\be
 L = - \frac{\pi\kappa}{\theta} \tr\left(
 -\epsilon_{ij}X^i\left(\dot{X}^j -i\left[A_0,X^j\right]\right)
 +2\theta A_0\right).
\ee
We may introduce the complex notation
\be
 c &=& \frac{1}{\sqrt{2\theta}}\left(x^1-ix^2\right),\nonumber\\
 K &=& \frac{1}{\sqrt{2\theta}}\left(X^1-iX^2\right)~,
\ee
in which the Lagrangian can be written \cite{Bak:2001sg}
\be
L = i\pi\kappa\tr\left(K^\dg D_t K - K D_t K^\dg\right) -2\pi\kappa
\tr A_0~,
\ee
where we have also introduced $D_t K = \partial_t K -i\left[A_0,K\right]$.
The equations of motion are
\be\label{eq:pureeom}
\left[K,K^\dg\right] &=& 1~,\nonumber\\
D_t K &=& 0~.
\ee
Susskind argues \cite{Susskind:2001fb} that in the presence of a
quasiparticle the first equation of (\ref{eq:pureeom}) gets modified to
\be
\left[K,K^\dg\right] &=& 1 + q|0\rangle\langle 0|,
\ee
where $\abs{q}<1$. In \cite{BOB} this system was studied from the
point of view of String Theory. The shape of the solutions was found
using the exact Seiberg-Witten map \cite{sw1,sw2,sw3,sw4} and it
was shown that the solution
represents a deficit or excess of charge centered around the
origin. However, this solution does not have any moduli, which we
need to study moving solutions. The situation improves when we include a scalar field, which has been
argued to describe density fluctuations of the quantum Hall liquid
\cite{Bak:2001sg}. In this case the solutions have moduli describing
the positions of the independent solitons, so we should be able to make
them move around.

\section{The nonrelativistic scalar field}\label{sec:nr}
The Lagrangian for noncommutative Chern-Simons theory coupled to a
nonrelativistic scalar field can be written as
\cite{Bak:2001sg,Jackiw:1990mb}
\be\label{eq:paramdef}
L = L_{CS} + 2\pi\theta\tr\left(iD_t\phi\phi^\dg - \frac{1}{2m}D_i\phi
(D_i\phi)^\dg+\lambda(\phi\phi^\dg)^2\right),
\ee
where $m$ is a mass parameter of dimension $({\rm length})^{-1}$ and
$\lambda$ is a parameter of dimension length.
This theory has a BPS bound in the case where $2m\lambda\kappa
=1$. Assuming this relation, rescaling the time variable\ $t\to mt$ and
introducing the complex notation
\be
D = \sqrt{\frac{\theta}{2}}\left(D_1-iD_2\right),
\ee
we may write the action in operator notation as\footnote{In fact, a further rescaling is possible which leaves an action with the only free parameter an overall factor of $\kappa$; $t\to \theta t,~\phi\to \sqrt{\kappa/\theta}\phi, ~A_0\to A_0/(m\theta)$.}
\be\label{niceact}
L&=&i\pi\kappa\tr\left\{K^{\dg}D_tK - KD_t{K}^{\dg}\right\} -
2\pi\kappa \tr A_0
\nonumber\\
&&+2\pi\theta\tr\left\{ i D_t \phi \phi^{\dg} - \frac{1}{2\theta}\left(
D\phi(D\phi)^{\dg} + \bar{D}\phi(\bar{D}\phi)^{\dg}\right)
+\frac{1}{2\kappa}(\phi\phi^{\dg})^2\right\},
\ee
where $D,\bar{D}$ and $D_t$ are defined as
\be
\label{definitions}
D_t\phi &=& \dot{\phi}-iA_0\phi,\nonumber\\
D\phi &=& \left(K\phi-\phi c\right),\\
\bar{D}\phi &=& -\left(K^{\dg}\phi - \phi c^{\dg}\right). \nonumber
\ee
The action (\ref{niceact}) is gauge invariant under the transformations
\be\label{gaugetransf}
\phi\to U\phi~,~~K\to U K U^\dg~,~~A_0\to U(i\partial_t+A_0)U^\dg~,
\ee
where $U$ is an arbitrary unitary operator. The action is also invariant under translations by a constant $z$:
\be\label{transtransf}
\phi\to\phi U(z)~,~~K \to K + z~,~~ A_0 \to A_0
~~;~~~U(z)\equiv e^{z c^\dg-\bar z c} ~.
\ee

Extremizing the action (\ref{niceact}) gives the equations of motion
\be
\label{EOM:nr}
iD_t\phi + \frac{1}{2\theta} \left\{D,\bar{D}\right\}\phi
+\frac{1}{\kappa}\phi\phi^{\dg}\phi &=& 0~,
\nonumber\\
1 - [K,K^{\dg}]  &=& \frac{\theta}{\kappa}\phi\phi^{\dg}~,
\\
\left\{K,\phi\phi^{\dg}\right\}-2\phi c \phi^{\dg} &=&
2\kappa i D_t K~.\nonumber
\ee
In the time-independent case one can show that any solution to the BPS
equations
\be
\label{BPS:nr}
K^{\dg}\phi - \phi c^{\dg} &=& 0~, \nonumber\\
1-\left[K,K^\dg\right] &=& \frac{\theta}{\kappa}\phi\phi^\dg~,\\
A_0 &=& -\frac{1}{2\kappa}\phi\phi^\dg~,\nonumber
\ee
is also a solution to the full equations of motion.

\subsection{The one soliton case}

\subsubsection{The solution}
The static one-soliton solution discussed in \cite{Bak:2001sg} looks like
\be
\label{static:simple}
\phi &=& \sqrt{\frac{\kappa}{\theta}}
\ket{0}\bra{z},
\nonumber\\
K &=& z \ket{0}\bra{0} + S_1 c S_1^{\dg},\\
A_0 &=& -\frac{1}{2\theta}\ket{0}\bra{0},\nonumber
\ee
where we have introduced the shift operators $S_n = \sum_{i=0}^{\infty}
\ket{i+n}\bra{i}$ and the usual coherent state $\ket{z} =
U(z)\ket{0}$. This solution can be found from the simple soliton at $z=0$ by a combination of the translation (\ref{transtransf}) and a gauge transformation with a unitary operator
\be\label{transgauge}
T(z)=e^{zK^\dg-\bar zK}~.
\ee

\subsubsection{The shape of the one soliton solution}
To find the shape of the soliton solution we have to extract it from
the operators $X_1$ and $X_2$, or equivalently from $K =
\frac{1}{\sqrt{2\theta}} \left(X^1 -iX^2\right)$. This we do by
using the exact Seiberg-Witten map \cite{sw1,sw2,sw3,sw4}, which implies that the Fourier transform
of the density is given by
\be\label{swmap}
 \tilde{\rho}({\bf k}) = \tr \left( e^{-ik_a X^{a}}\right) =
 \tr \left( e^{-i(\bar{k}K + k K^\dg)}\right)~,
\ee
where we have defined $k =\sqrt{\frac{\theta}{2}}(k_1 - i k_2)$.
We then have
\be
\rho({\bf x}) = \int \frac{d^2 {\bf k}}{(2\pi)^2} \tilde{\rho}({\bf k})
e^{ik_a x^a}
\ee
where $x^a$ are ordinary commuting coordinates. This form of
the density is very natural if one thinks of the coordinates of the individual electrons as the eigenvalues of the $X$ matrices, since then $\tilde{\rho} =
\sum_{i=1}^{N}e^{-i{\bf k\cdot x}_{i}}$ is exactly the Fourier
transform of the density $\rho = \sum_{i=1}^{N} \delta({\bf x - x}_i)$.

To evaluate the Fourier moments we first notice that $K$ is composed
of two pieces which commute with each other. This makes it possible to
write
\be
\tr \left(e^{-i(\bar{k}K + k K^\dg)}\right) =
\str \left(e^{-i(\bar{k}z+k\bar{z})\ket{0}\bra{0}}\right) +
\tr' \left(e^{-i(\bar{k}S_1cS_1^\dg + k S_1c^\dg S_1^\dg)}\right)
\ee
where we have introduced the notation $\str$ for traces over a finite
set of states (in this case over the set $\ket{0}$) and $\tr'$ for an
infinite trace with some states omitted (in this case $\ket{0}$ is
omitted). These traces can be evaluated separately. The first one is
\be
\bra{0}\left(e^{-i(\bar{k}z+k\bar{z})\ket{0}\bra{0}}\right)\ket{0} =
e^{-i(\bar{k}z+k\bar{z})}
\ee
and the second one is
\be
\sum_{n=1}^{\infty} \bra{n}\left(e^{-i(\bar{k}S_1cS_1^\dg + k S_1c^\dg
S_1^\dg)}\right) \ket{n} =
\sum_{n=1}^{\infty} \bra{n}\left(S_1 e^{-i(\bar{k}c+kc^\dg)}S_1^\dg
+1-S_1S_1^\dg\right)\ket{n}
\ee
Noticing that $S_1^\dg$ will shift the state $\ket{n}$ to $\ket{n-1}$
and that $1-S_1S_1^\dg$ is the projection operator on the space
spanned by $\ket{0}$, we get
\be
\tr' \left(e^{-i(\bar{k}S_1cS_1^\dg + k S_1c^\dg S_1^\dg)}\right) =
\tr \left(e^{-i(\bar{k}c+kc^\dg)}\right)
\ee
This trace we perform using coherent states
\be
\tr \left(e^{-i(\bar{k}c+kc^\dg)}\right) &=& \int\frac{d^2 \rm w}{\pi} \bra{w}
e^{-i(\bar{k}c+kc^\dg)}\ket{w}
\nonumber\\
&=& \int\frac{d^2 \rm w}{\pi} \bra{w}
e^{-ikc^\dg}e^{-i\bar{k}c}\ket{w} \;e^{-\frac{k\bar{k}}{2}} \\
& =&\int\frac{d^2 \rm w}{\pi} e^{-i(k\bar{\rm w}+\bar{k}{\rm
w})}e^{-\frac{k\bar{k}}{2}} \nonumber\\
&= &\frac{2\pi}{\theta}\delta^{(2)}({\bf k})~.\nonumber
\ee
Thus we find the result
\be
\tr \left(e^{-i(\bar{k}K+kK^\dg)}\right) =
\frac{2\pi}{\theta}\delta^{(2)}({\bf k}) + e^{-i(\bar{k}z+k\bar{z})}~.
\ee
Fourier transforming back we get
\be
\rho({\bf x}) = \frac{1}{2\pi\theta} +
\delta(x_1-\sqrt{2\theta}\Re(z))
\delta(x_2-\sqrt{2\theta}\Im(z))~,
\label{rho}
\ee
which is clearly to be interpreted as a constant background density
$\frac{1}{2\pi\theta}$ with a delta function type deformation\footnote{Note that the sign of the deformation is opposite to that of the quasiparticles studied by Susskind \cite{Susskind:2001fb}.} located
at $\sqrt{2\theta}z$.

\subsubsection{More stationary solutions}
The configuration (\ref{static:simple}) turns out to be a member of a huge
family of stationary solitons with a time independent density of the form
(\ref{rho}). To see this consider an Ansatz
\begin{equation}
\label{ansatz1}
\phi = \sqrt{\frac{\kappa}{\theta}}\op{0}{\psi},
\hskip 1cm
A_0 = -\frac{1}{2\theta} \op{0}{0},
\hskip 1cm
K = S_1 c S_1^\dag.
\end{equation}
Inserting (\ref{ansatz1}) into the equations of motion (\ref{EOM:nr}) we get
\begin{eqnarray}
\label{eq11}
\ip{\psi}{\psi} & = & 1,\\
\label{eq21}
\bra{\psi} c \ket {\psi} & = & 0, \\
\label{eq31}
-i\partial_{\tau}\ket{\psi} & = & c^\dag c\ \ket{\psi},
\end{eqnarray}
where $\tau = {t\over \theta}.$

The general solution of (\ref{eq31}) reads
\begin{equation}
\label{solution}
\ket{\psi} = \sum\limits_{n=0}^{\infty}
\frac{b_n {\rm e}^{in\tau}}{\sqrt{n!}}\ket{n}
\end{equation}
where $b_n$ are time-independent and satisfy
\begin{eqnarray}
\label{eq12}
\sum\limits_{n=0}^{\infty}\frac{|b_n|^2}{n!} & = & 1, \\
\label{eq22}
\sum\limits_{n=0}^{\infty}\frac{\overline{b_n}b_{n+1}}{n!} & = & 0.
\end{eqnarray}

If we introduce the generating function
\[
G(\zeta) =\sum\limits_{n=0}^{\infty}\frac{b_n}{n!}\zeta^n
\]
then (\ref{eq12}) and (\ref{eq22}) can be rewritten as
\begin{eqnarray}
\label{eq13}
\frac{1}{\pi}\int\limits_{\mathbb C}d^2\zeta\ \overline{G(\zeta)} G(\zeta)
{\rm e}^{-\zeta\bar\zeta}
& = & 1, \\
\label{eq23}
\frac{1}{\pi}\int\limits_{\mathbb C}d^2\zeta\
\overline{\zeta\,G(\zeta)} G(\zeta) 
{\rm e}^{-\zeta\bar\zeta}
& = & 0.
\end{eqnarray}
Any function $G$ which satisfies these relations yields a solution to
(\ref{EOM:nr}). 

We may think of the states $\ket{\psi_n}\equiv e^{in\tau}\ket{n}$ as a basis and the
combination (\ref{solution}) as the expression for a general state in
this basis. The energy of a configuration can be computed from
\be
E = \pi\tr\left(D\phi\left(D\phi\right)^\dg +
\bar{D}\phi\left(\bar{D}\phi\right)^\dg -
\frac{\theta}{\kappa}\left(\phi\phi^\dg\right)^2\right).
\ee
which for our particular ansatz becomes
\be
E= \frac{\pi\kappa}{\theta}\left(
\bra{\psi}cc^\dg\ket{\psi} + \bra{\psi}c^\dg c\ket{\psi}
-\ip{\psi}{\psi}\right) = 
\frac{2\pi\kappa}{\theta}\bra{\psi}c^\dg c\ket{\psi}.
\ee
The ``basis'' states thus have energy
\be
E_n = \frac{2\pi\kappa}{\theta} n,
\ee
and a general state has energy
\be
E &  = &
\frac{2\pi\kappa}{\theta} \sum\limits_{n=1}^{\infty}\frac{|b_n|^2}{(n-1)!}
=
\frac{2\kappa}{\theta}\int\limits_{\mathbb C}d^2\zeta\ \left|\zeta\,G(\zeta)\right|^2
{\rm e}^{-\zeta\bar\zeta}~,
\ee
which can be arbitrary.

More general solutions can be found by translating these solitons to arbitrary $z$ using the transformations (\ref{transtransf}); a simpler form of the translated solution can be found by combining the translation with the gauge transformation (\ref{transgauge}). The naive energy {\em density} is not gauge invariant; the physical energy density is found by performing a Seiberg-Witten map (\ref{swmap}). We find that the physical energy density is translated and is concentrated at the position of the soliton.

The basis states $\ket{\psi_n}$ can be distinguished from more general configurations in a gauge invariant way. If we consider invariants of the form 
\be
I[f]=\tr(\phi^\dg\phi f(c,c^\dg))~,
\ee
then for generic functions $f$, we find that $I[f]$ is time independent only for the basis states. The original soliton (\ref{static:simple}) is the only solution
to the BPS equations (\ref{BPS:nr}) in the family constructed here.

\subsubsection{Moving one-soliton solutions}
To make the solitons move one may use the ``exotic'' Galilean
invariance of the theory. The infinitesimal version of this
tranformation was found in \cite{galil}. 
Defining an infinitesimal complex paramteter
$v=\frac{1}{\sqrt{2\theta}}(v_1-iv_2)$ and writing everything in the
operator notation, the transformations look like
\be
\delta\phi &=& t\left[vc^\dg-\bar{v} c,\phi\right] + i\theta \phi
\left(vc^\dg+\bar{v} c\right)~,\nonumber\\
\delta K &=& t\left[vc^\dg-\bar{v} c,K\right]+vt~,\\
\delta A_0 &=& t\left[vc^\dg-\bar{v} c,A_0\right]
+i\left(v(K^\dg-c^\dg)-\bar{v}(K-c)\right)~.\nonumber
\ee
This transformation can be integrated to finite values of the
parameter $v$. The transformation becomes
\be\label{galtransf}
\phi &\to& U\phi V^\dg~,\nonumber\\
K &\to& U(K+vt)U^\dg~,\\
A_0 &\to& U\left(A_0+i\Big(v(K^\dg-c^\dg)-\bar{v}(K-c)\Big)\right)U^\dg~,
\nonumber
\ee
which can be checked to leave the equations of motion invariant. Here $U$ is the unitary translation operator which translates a state by $vt$
\be
U(vt) = e^{(vc^\dg-\bar{v} c)t}~,
\ee
and $V$ is also a unitary translation operator, but it translates by $v(t-i\theta)$. 
It is given by
\be
V(vt) = e^{v(t-i\theta)c^\dg-\bar{v}(t+i\theta)c}~.
\ee
The transformations (\ref{galtransf}) close under group multiplication up to a U(1) gauge transformation:
\be
U(v_1t)U(v_2t)\phi V^\dg(v_2t) V^\dg(v_1t)=
e^{\frac12\theta^2(v_2\bar v_1- v_1 \bar v_2)}
U((v_1+v_2)t)\phi V^\dg((v_1+v_2)t)~;
\ee
as the U(1) transformation is constant and commutative, it does not act on $K$ and $A_0$.

The form of the Galilean tranformation can be improved by acting with
an additional gauge transformation (\ref{gaugetransf}). In particular, choosing
the unitary operator $U^\dg(vt)$, one can write the action of the Galilean transformation as
\be
\phi &\to& \phi V^\dg~,\nonumber\\
K &\to& K +vt~,\\
A_0 &\to& A_0 +i\left(v K^\dg - \bar{v} K\right)~.
\nonumber
\ee
Note the similarity to the translations (\ref{transtransf}).

When $K$ is time-independent, the shift in $A_0$ can then be eliminated by a further gauge transformation with 
\be
T = e^{tvK^\dg-t\bar{v} K}~,
\ee
giving us the final galilean transformation 
\be\label{boosttransf}
\phi &\to& T\phi V^\dg~,\nonumber\\
K &\to& T K T^\dg +vt~,\\
A_0 &\to& T A_0 T^\dg~;\nonumber
\ee
we emphasize that the simple homogeneous transformation of $A_0$ requires 
time-independent $K$.

As an example, the
$z_0=0$ static soliton
\be
 \phi &=& \sqrt{\frac{\kappa}{\theta}}\ket{0}\bra{0}~,\nonumber\\
 K &=& S_1 c S_1^\dg~,\\
 A_0 &=& -\frac{1}{2\theta}\ket{0}\bra{0}~,\nonumber
\ee
can be boosted using (\ref{boosttransf}), as $K$ is time-independent.
Here, the operator $T$ is block diagonal and acts
trivially on $\ket{0}$. Also, one may show that $TKT^\dg = K
-vt\left[K,K^\dg\right]$. Thus we find the boosted operators
\be
 \phi &=& \sqrt{\frac{\kappa}{\theta}}\ket{0}\bra{v(t-i\theta)}~,
\nonumber\\
 K &=& vt\ket{0}\bra{0} + S_1 c S_1^\dg~,\\
 A_0 &=& -\frac{1}{2\theta}\ket{0}\bra{0}~,\nonumber
\ee
where $\ket{v(t-i\theta)}$ is the coherent state at $z=v(t-i\theta)$. Other solutions can be generated by boosting more general stationary solutions, {\it e.g.}, solutions centered at some point $z_0$ or the new stationary solutions described above.

\subsection{The two soliton case}

\subsubsection{The solution}
After performing the gauge transformation (\ref{gaugetransf}) with
\(
U = \op{0}{1} + \op{1}{0} +  S_2  S_2^\dg
\)
the two soliton solution found in \cite{Bak:2001sg}
can be written as
\be
\label{2sol:nrel}
\phi &=& \sqrt{\frac{2\kappa}{\theta}}
\left(A\ket{0}\bra{+} + B \ket{1}\bra{-}\right)\\
K &=& z \left(C \ket{0}\bra{1} + \frac{1}{C} \ket{1}\bra{0}\right)
+ S_2 c S_2^\dg,
\nonumber
\ee
where $\ket{\pm}$ are the normalized states
\[
\ket{\pm} = \frac{1}{\sqrt{2\left(1 \pm {\rm e}^{-2|z|^2}\right)}}
\left(\ket{z} \pm \ket{-z}\right),
\hskip 1cm
\ket{z} = {\rm e}^{zc^\dg - \bar z c}\ket{0}.
\]
The BPS equations now tell us that $\abs{A}^2 + \abs{B}^2 = 1$, which
is solved by $A = \sin\left(\alpha\right)$ and $B =
\cos\left(\alpha\right)$, that $C = \cot\left(\alpha\right)
\sqrt{\coth\abs{z}^2}$, and that $\alpha$ has to satisfy the
equation
\be
\abs{z}^2\left(C^2-\frac{1}{C^2}\right) = B^2 - A^2~.
\ee
This is a cubic equation in $\tan\alpha$, so it can be
solved explicitly. The solution is not particularly illuminating, so we
explicitly give only some limiting
properties and use the full result in the calculations. We find that
\be\label{eq:asyC}
C &=& 1-\frac{e^{-2\abs{z}^2}}{4\abs{z}^2-1}+
 {\cal O}\left(e^{-4\abs{z}^2}\right) \;\; \abs{z}\gg 1
\nonumber\\
C &=&\abs{z}\left(1+\frac{1}{2}\abs{z}^4-\frac{11}{24}\abs{z}^8+
{\cal O}\left(\abs{z}^{12}\right)\right) \;\; \abs{z} \ll 1~.
\ee

\subsubsection{The shape of the two soliton solution}
\label{the:shape}
Using the same method as in the one soliton case we can calculate the
shape of the two soliton solution. The Fourier moments of the density
are given by
\be
\tilde{\rho}({\bf k}) = \tr \left( e^{-i(\bar{k}K+kK^\dg)} \right)
\ee
which again splits into two pieces
\be
\tr \left( e^{-i(\bar{k}K+kK^\dg)} \right) =
 \str\left(e^{-i(D\ket{0}\bra{1}+\bar{D}\ket{1}\bra{0})}\right) +
 \tr' \left(e^{-i(\bar{k}S_2cS_2^\dg+kS_2c^\dg S_2^\dg)}\right)
\ee
where we have introduced $D=\bar{k}zC+\frac{k\bar{z}}{C}$ and where
$\str$ now means the trace over the subspace spanned by
$\ket{0},\ket{1}$ and $\tr'$ means the trace over the orthogonal
complement of this space. Using the methods developed for the one
soliton case we can evaluate this as
\be
\tr' \left(e^{-i(\bar{k}S_2cS_2^\dg+kS_2c^\dg S_2^\dg)}\right) =
\tr \left(e^{-i(\bar{k}c+kc^\dg)}\right)~,
\ee
which again leads to a constant background density, and
\be
\str\left(e^{-i(D\ket{0}\bra{1}+\bar{D}\ket{1}\bra{0})}\right) =
2 \cos\left(\abs{D}\right)~,
\ee
which gives us the perturbation part of the density
\be
\rho_p({\bf x}) = \int \frac{d^2 {\bf k}}{(2\pi)^2}
2\cos\left(\abs{D}\right) e^{ik_ax^a}~.
\ee
To do this integral we choose integration
variables $k$ that simplify $\abs{D}$: we change
coordinates from ${\bf k}$ to ${\bf q}$
\be
k = \frac{z}{2\abs{z}}\left(\frac{q_1}{\lambda_+}+i
\frac{q_2}{\lambda_-}\right) ~,
\ee
where we have defined
\be
\lambda_{\pm} = \frac{\abs{z}}{2}\left(\frac{1}{C}\pm C\right).
\ee
In these coordinates $D$ is simply
\be
\abs{D}^2 =q_1^2 + q_2^2~.
\ee
Using the asymptotic expressions for C
(\ref{eq:asyC}) we see that
\be
\lambda_+ &=& \left\{\begin{array}{ccc}
\frac{1}{2} & ~~~ & \abs{z}\ll 1\\
\abs{z} & ~~~ & \abs{z} \gg 1
\end{array}\right.
\nonumber\\
\lambda_- &=& \left\{\begin{array}{ccc}
\,\frac{1}{2}\, & ~~~ & \abs{z} \ll 1\\
\,0\, &  ~~~ & \abs{z} \gg 1
\end{array}\right.
\ee
The Jacobian is $\frac{1}{2\theta\lambda_+\lambda_-}$ so we have
\be
\rho_p({\bf y}) = \frac{1}{\theta\lambda_+\lambda_-}
 \int \frac{d^2{\bf q}}{(2\pi)^2}
\cos\left(\sqrt{q_1^2 + q_2^2}\right) e^{iq_a y^a}
\ee
where we have introduced a new variable ${\bf y}$ defined by ${\bf k
  \cdot x \equiv q \cdot y}$, which implies
\be
(y_1,y_2)=\left(\frac1{4\abs{z}\lambda_+}[x_1(z+\bar z)+ix_2(\bar z - z)]~,~
\frac1{4\abs{z}\lambda_-}[x_2(z+\bar z)-ix_1(\bar z - z)] \right).
\ee
Note that a circle of radius $1$ in the ${\bf y}$ coordinate system when
transformed to ${\bf x}$ coordinates looks like an ellipse
with the major axis oriented in the direction of $z$ and of length
$\sqrt{2\theta}\lambda_+$ and a minor axis of length
$\sqrt{2\theta}\lambda_-$. From (\ref{eq:asyC}) we also see
that for small $\abs{z}$ the ellipse becomes a circle of radius
$\sqrt{\frac{\theta}{2}}$, whereas for large $\abs{z}$, $\lambda_-$ goes exponentially fast to zero so the ellipse is rapidly squeezed into a line between $\pm\sqrt{2\theta}\lambda_+ \rightarrow \pm \sqrt{2\theta}\abs{z}$.

We perform the angular integral and find
\be
\rho_p({\bf y}) = \frac{1}{\theta\lambda_+\lambda_-}
\int \frac{dq}{2\pi} \:q\: \cos q \:J_0(qy) =
\frac{1}{\theta\lambda_+\lambda_-} \Re
\int \frac{dq}{2\pi} \:q\: e^{-iq} \:J_0(qy)
\ee
where $y$ is the absolute value of the vector ${\bf y}$. This integral
is divergent and needs to be regulated. This we do by adding a small
real part to the argument in the exponential. Thus, with $a=i+\epsilon$ we
have
\be
\rho_p({\bf y}) = \lim_{\epsilon\rightarrow 0}
\frac{1}{\theta\lambda_+\lambda_-} \Re
\int \frac{dq}{2\pi} \:q\: e^{-aq} \:J_0(qy) =
\lim_{\epsilon\rightarrow 0}
\frac{1}{2\pi\theta\lambda_+\lambda_-} \Re
\frac{a}{\left(y^2+a^2\right)^{\frac{3}{2}}}
\ee
which is completely well defined and
smooth for all non-zero $\epsilon$. Letting
$\epsilon\rightarrow 0$ the function is zero for $y>1$
and for $y<1$ it looks like $\rho_p = -\frac{1}{2\pi\theta\lambda_+\lambda_-}
\frac{1}{(1-y^2)^{\frac{3}{2}}}$. In figure \ref{fig:dens} we see a cross
section of the full density drawn in the $y$-plane where the border of
the ellipse is
at $\abs{y}=1$ and for a small non-zero value of $\epsilon$. The
background density is normalized to one.
\FIGURE[v]{
\mbox{\epsfxsize=10cm\epsfbox{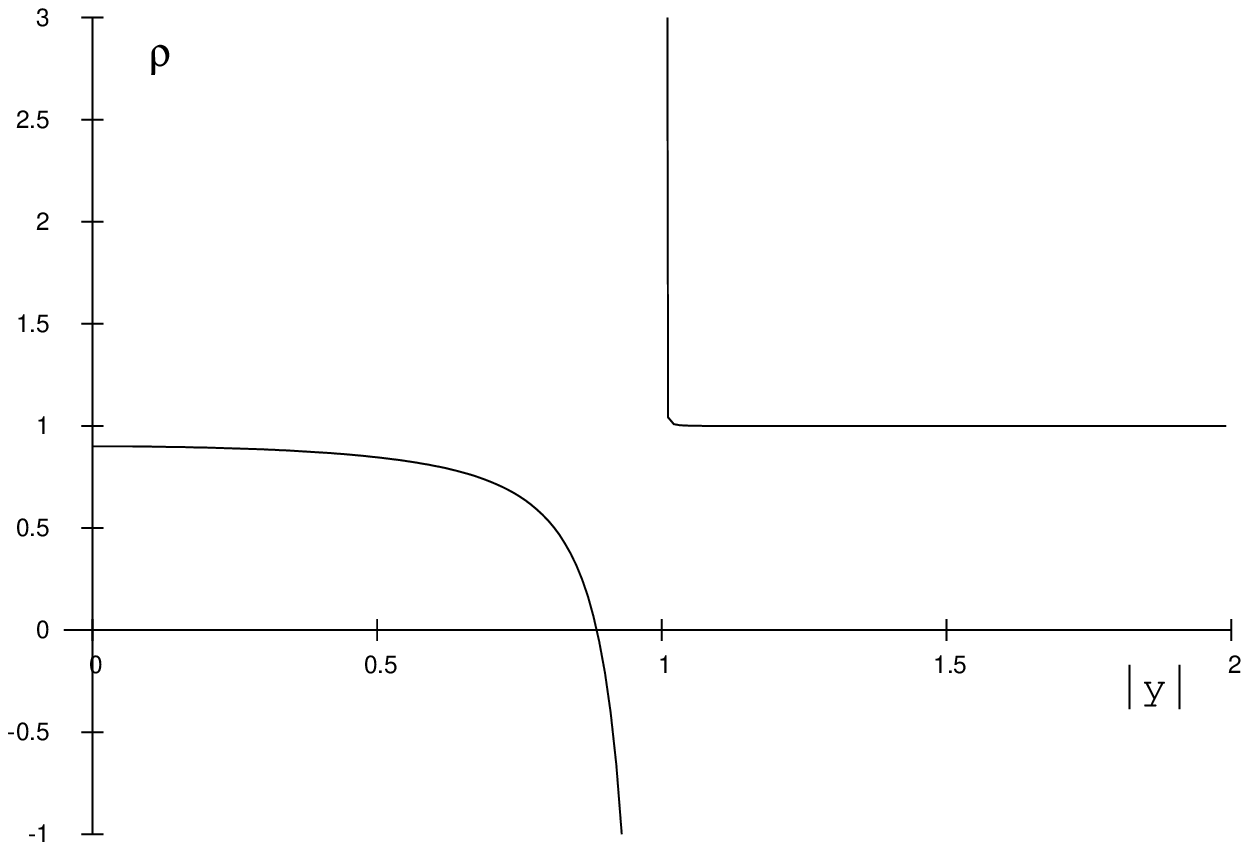}}
\caption{The full density\label{fig:dens}}
}

Notice that the full density inside the ellipse is always less than the
background density. This
means that we have taken particles from the inside of the ellipse and
piled them up at the border of the ellipse. Notice also that there is
always a small region on the inside of the border
of the ellipse where the full density will be negative. This
seems difficult to reconcile with the interpretation that the theory
describes an electron gas since then there could not be less than zero
electrons at any point.

As a final exercise we show how two one-soliton solution can be found
in the $\abs{z}\rightarrow \infty$ limit of the two-soliton
solution. We want to show that
\be
\lim_{\abs{z}\rightarrow\infty} \rho_p({\bf x}) =
\delta^{(2)}\left(x-\sqrt{2\theta}z\right) +
\delta^{(2)}\left(x+\sqrt{2\theta}z\right)
\ee
To do this we have to use the regularized version of $\rho_p$
using the non-zero $\epsilon$ which we will take to zero only at the
end of the calculation. Thus we would like to show that for any
real function $f(x_1,x_2)$
\be
f(\sqrt{2\theta}z)+f(-\sqrt{2\theta}z) &=& \lim_{\epsilon\rightarrow 0}
\lim_{\abs{z}\rightarrow\infty}\int d^{2}{\bf x} f({\bf x})
\rho_p({\bf x}) \nonumber\\
&=&\lim_{\epsilon\rightarrow 0} \lim_{\abs{z}\rightarrow\infty}
\frac{1}{2\pi\theta\lambda_+\lambda_-}
\Re \int d^{2}{\bf x} f({\bf x})
\frac{i}{(y(x)^2+(i+\epsilon)^2)^{\frac{3}{2}}}
\ee
Changing integration variables from ${\bf x}$ to ${\bf y}$, not
forgetting the Jacobian, we get
\be
\lim_{\epsilon\rightarrow 0} \lim_{\abs{z}\rightarrow\infty}
\frac{1}{\pi}
\Re \int d^{2}{\bf y}\; f(\sqrt{2\theta}\lambda_+
y_1,\sqrt{2\theta}\lambda_- y_2) \;
\frac{i}{(y^2+(i+\epsilon)^2)^{\frac{3}{2}}}
\ee
But in the limit when $\abs{z}$ goes to infinity, $\lambda_-$ goes to
zero exponentially fast. This means that the $y_2$ dependence of the
function $f$ is scaled out and the integral becomes
\be
\lim_{\epsilon\rightarrow 0} \lim_{\abs{z}\rightarrow\infty}
\frac{1}{\pi}
\Re \int dy_1\; f(\sqrt{2\theta}\lambda_+
y_1,0)  \int_{-\infty}^{+\infty} dy_2
\frac{i}{(y^2+(i+\epsilon)^2)^{\frac{3}{2}}}
\nonumber\\
=\lim_{\epsilon\rightarrow 0} \lim_{\abs{z}\rightarrow\infty}
\frac{1}{\pi}
\Re \int dy_1\; f(\sqrt{2\theta}\lambda_+
y_1,0) \frac{2i}{y_1^2+(i+\epsilon)^2}
\ee
Taking the real part and performing the $\epsilon\to 0$ limit we recognize
a representation of the delta function
\be
\lim_{\epsilon\rightarrow 0} \Re
\frac{2i}{y_1^2-1+i\epsilon} = \lim_{\epsilon\rightarrow 0}
\frac{2\epsilon}{(y_1^2-1)^2+\epsilon^2} = 2\pi\delta(y_1^2-1) =
\pi\delta(y_1-1)+\pi\delta(y_1+1)
\ee
and substituting this back we get
\be
f(\sqrt{2\theta}\lambda_+,0)+f(-\sqrt{2\theta}\lambda_+,0)
\ee
and since for large $\abs{z}$ we have $\lambda_+ \approx \abs{z}$,
this is what we wanted to show.

\subsubsection{The dynamics}
To study the dynamics on the two soliton moduli space one can
try to make the moduli slowly time dependent and find the effective
action in the moduli. Doing this we find
\be
i\pi\kappa\int
dt\left(C^2-\frac{1}{C^2}\right)^2
\abs{z}^2(\bar{z}\dot{z}-z\dot{\bar{z}})~,
\ee
or, using $z=re^{i\phi}$,
\be
-2\pi\kappa \int dt \left(C^2-\frac{1}{C^2}\right)^2r^4\dot{\phi}~.
\ee
Notice that there is no time derivative of $r$ and that the function
in front of $\dot{\phi}$ depends only on $r$. It can
be written as $(4\lambda_+\lambda_-)^2$ which is identical to the
``angular momentum'' of the two soliton configuration defined in
\cite{Bak:2001sg}. Using the asymptotic expansions for C
(\ref{eq:asyC}) we find that it goes as $(1-4r^4)$ for small $r$ and
for large $r$ it goes exponentially to zero.

The effective action is first order in time derivatives and thus comes
in ``Hamiltonian'' form (with the Hamiltonian equal to zero). It
defines a connection 1-form on the moduli
space ${\cal A} =
i\left(C^2-\frac{1}{C^2}\right)^2\abs{z}^2\left(zd\bar{z}-\bar{z}dz\right)$
whose curvature 2-form is the symplectic form on the moduli space seen
as a phase space. This may be taken as the starting point for
quantization of the system but since the Hamiltonian is zero, there is
no non-trivial time evolution. We have plotted the symplectic form in
figure \ref{fig:symp}.
\FIGURE[v]{
\mbox{\epsfxsize=10cm\epsfbox{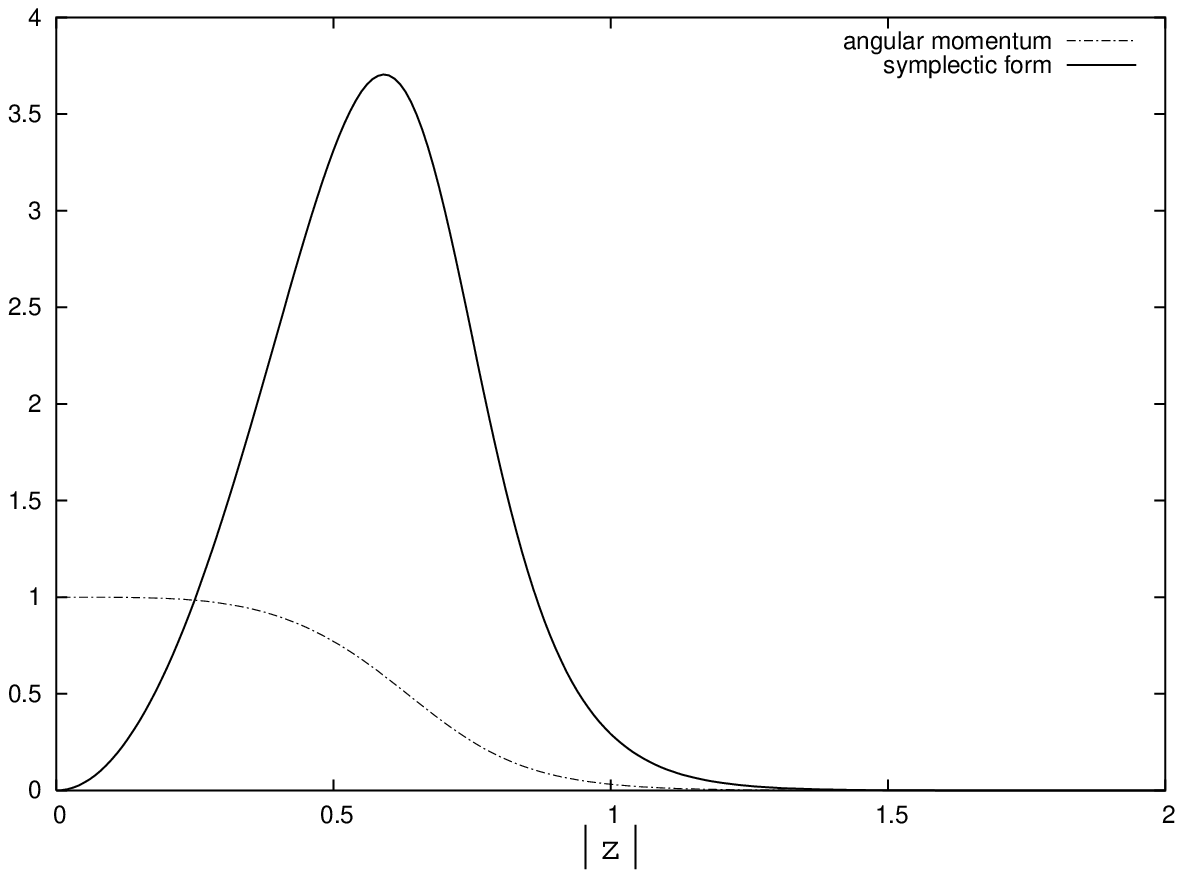}}
\caption{The angular momentum and the symplectic
2-form\label{fig:symp}}}
At small $\abs{z}$ it goes to zero as $16\abs{z}^2$ and at
large $\abs{z}$ it goes to zero exponentially\footnote{Since the form degenerates at $0$ and $\infty$, the phase space is singular at these points.}. In between (at $\abs{z}\approx 0.6$) there is a maximum.

One could speculate that, as in \cite{Manton:1997tg}, a potential for
the moduli could be generated. This would then define a nontrivial
Hamiltonian which would give non-trivial dynamics on the moduli space.

As found in \cite{Manton:1997tg,Romao:2000ru,Ulf} in the commutative
Chern-Simons case, the symplectic form defined above equals the
K\"ahler form of the metric on the moduli space of vortex solutions to
the Maxwell-Higgs model. If this is true also in the non-commutative
model, the symplectic form we just computed should have the properties
of a K\"ahler metric. This seems to be true for small $\abs{z}$ where
the apparent conical singularity is removed after dividing by the
$Z_2$ action that arises because the voritces are identical
particles. At large $\abs{z}$ however, the metric is zero, which seems
to go against the intuition that the metric should be flat for well
separated vortices (they should not interact at large
distances). It also seems to disagree with explicit calculations of the
metric in the Maxwell-Higgs case \cite{Tong:2002xt}. However, our
result is similar to what we find in the
relativistic case in the next section, where we again find a nontrivial
metric between vortices even in the large $\abs{z}$ limit. As in
\cite{Ulf} one should be able to use our expression for the symplectic
two form to calculate the phase space volume or rather the reduction
of the phase space volume due to the presence of other vortices. From
this one could find the statistics of the vortices.

There have been other attempts to find the
dynamics of first order system of vortices, e.g.,
\cite{Dziarmaga:1994yn}. However, without additional assumptions the
methods of \cite{Dziarmaga:1994yn} do not yield unique solutions, and
we were not able to put them to fruitful use in our case.

\section{ The relativistic scalar field}
The Lagrangian of a noncommutative Chern-Simons field interacting with
a {\em relativistic} scalar field can be written \cite{Bak:2001sg}
(using the objects defined in (\ref{definitions})) as
\be
\label{lagrangian:rel}
{\rm L}_{\rm rel} & = & i\pi\kappa\tr\left\{K^{\dg}D_tK -
KD_t{K}^{\dg}\right\} -  2\pi\kappa \tr A_0
+\nonumber\\
&& +2\pi\theta\tr\left\{ D_t \phi \left(D_t\phi\right)^{\dg}
 - \frac{1}{\theta}\left(
D\phi(D\phi)^{\dg} + \bar{D}\phi(\bar{D}\phi)^{\dg}\right)
-V(\phi\phi^{\dg})\right\}~,
\ee
leading to the equations of motion
\be
\label{EOM:rel1}
D_tD_t\phi - \frac{1}{\theta} \left\{D,\bar{D}\right\}\phi
+V'(\phi\phi^{\dg})\phi =  0 ~,
\ee
\be
\label{EOM:rel2}
\kappa i D_t K  =
\left\{K,\phi\phi^{\dg}\right\}-2\phi c \phi^{\dg}~,
\ee
\be
\label{EOM:rel3}
1 - [K,K^{\dg}] =
i\frac{\theta}{\kappa}\left[D_t\phi\phi^{\dg} -
\phi\left(D_t\phi\right)^\dg\right] ~.
\ee
If the potential has the form
\begin{equation}
\label{potential}
V(\zeta) = \frac{1}{\kappa^2}\zeta(\zeta - u^2)^2~,
\end{equation}
where $u$ is a (real) constant, then the model has a Bogomolny
bound and BPS equations:
\be
\label{BPS:rel1}
\bar D\phi = 0~,
\ee
\be
\label{BPS:rel2}
D_t\phi = \frac{i}{\kappa}\left(\phi\phi^\dg-u^2\right)\phi ~,
\ee
\be
\label{BPS:rel3}
1 - [K,K^{\dg}] =
i\frac{\theta}{\kappa}\left[D_t\phi\phi^{\dg} -
\phi\left(D_t\phi\right)^\dg\right]~.
\ee
There is also a second set of BPS equations in this model, with $D\phi=0$
and $D_t\phi = -\frac{i}{\kappa}\left(\phi\phi^\dg-u^2\right)\phi$, but these
equations do not seem to have localized solutions as there are no
normalizable eigenvectors of the creation operator $c^\dg$.

It is instructive to see how (\ref{EOM:rel1},\ref{EOM:rel2},\ref{EOM:rel3}) follow from (\ref{BPS:rel1},\ref{BPS:rel2},\ref{BPS:rel3}). Substituting (\ref{BPS:rel2}) into (\ref{BPS:rel3}) we get
\[
1 - [K,K^{\dg}] =
- \frac{2\theta}{\kappa^2}\left(\phi\phi^{\dg}-u^2\right)\phi\phi^{\dg}~.
\]
{}From $\bar D\phi = 0$ and the definitions of $D, \bar D$ it follows that
\begin{eqnarray*}
-\frac{1}{\theta}\{D,\bar D\}\phi & = & \frac{1}{\theta}[D,\bar D]\phi
 = \frac{1}{\theta}\left(1 - [K,K^{\dg}]\right)\phi
=  -\frac{2}{\kappa^2}\left(\phi\phi^{\dg}-u^2\right)\phi\phi^{\dg}\phi~.
\end{eqnarray*}
Computing the $D_t$ derivative of (\ref{BPS:rel2}), we get
\be
D_tD_t\phi = -\frac{i}{\kappa}\left(u^2D_t\phi - D_t\phi\,\phi^\dg\phi
- \phi(D_t\phi)^\dg\phi - \phi\phi^\dg D_t\phi\right) ~;
\ee
hence
\be
D_tD_t\phi =-\frac{1}{\kappa^2}\left(\phi\phi^\dg-u^2\right)^2\phi~,
\ee
and thus
\[
D_tD_t\phi - \frac{1}{\theta}\{D,\bar D\}\phi
= -\frac{1}{\kappa^2}\left(\phi\phi^\dg - u^2\right)
\left(3\phi\phi^\dg - u^2\right)\phi \equiv -V'\left(\phi\phi^\dg\right)\phi~.
\]
For $V$ of the form (\ref{potential}), this is (\ref{EOM:rel1}).

To derive (\ref{EOM:rel2}), we write
\[
0 = D_t(\bar D\phi) = -(D_t K)^\dg\phi -K^\dg D_t\phi + D_t\phi c^\dg~,
\]
which gives
\be
\phi^\dg D_t K = c(D_t\phi)^\dg - (D_t\phi)^\dg K
=
-\frac{i}{\kappa}\left(c\phi^\dg\phi\phi^\dg-  \phi^\dg\phi\phi^\dg
K\right)~;
\ee
multiplying with $\phi$ on the left, we get
\[
\phi\phi^\dg\ i\kappa D_tK
=
\phi c\phi^\dg\phi\phi^\dg - (\phi\phi^\dg)^2 K~.
\]
Noting that $\bar D\phi = 0 \; \Rightarrow \; \phi\phi^\dg K =
\phi c\phi^\dg,$ this gives
\[
\phi\phi^\dg\ i\kappa D_tK
=  \phi\phi^\dg\left[K,\phi\phi^\dg\right]
= \phi\phi^\dg\left(\{K,\phi\phi^\dg\}-2\phi c\phi^\dg\right)~.
\]
Here we see that, unlike in the non-relativistic case, a solution to the
BPS equations imply a solution to the {\em full} equations of motion
and not just to the time independent equations of motion. This is
almost true. There could also exist BPS solutions where the operator
\[
i\kappa D_tK -\{K,\phi\phi^\dg\} +2\phi c\phi^\dg,
\]
is non-zero, but lies in the subspace orthogonal to the operator
$\phi\phi^\dg$. These ``solutions'' would then be solutions to the BPS
equations {\em only}.

\subsection{The one soliton case}

\subsubsection{The static solution}

The static, one soliton solution is given by the Ansatz \cite{Bak:2001sg}
\[
K = z \op{0}{0} + S_1cS_1^\dg,
\hskip 1cm
\phi = \lambda \op{0}{z},
\hskip 1cm
A_0 = \frac{b}{\sqrt{2\theta}}\op{0}{0}
\]
where the parameters $\lambda$ and $b$ are real and are determined in terms of the parameters $\kappa,\theta,u$ in the Lagrangian. It follows from the equations of motion (\ref{EOM:rel1},\ref{EOM:rel2},\ref{EOM:rel3}) that $b = \frac{\kappa}{\sqrt{2\theta}\lambda^2}$ and
\[
\frac{\lambda^2}{u^2}
=
\frac12\left(1 \pm \sqrt{1-\frac{2\kappa^2}{\theta u^4}}\right)
\hskip 5mm {\rm or}
\hskip 5mm
\frac{\lambda^2}{u^2}
=
\frac16\left(1 \pm \sqrt{1+\frac{6\kappa^2}{\theta u^4}}\right).
\]
The first two values correspond to configurations
that are also solutions in the Bogomol'nyi limit. Note that they exist
only for $\theta > 2\kappa^2/u^4$, whereas the remaining two solutions are well defined for any nonvanishing noncommutativity parameter.

The $K$ part of the solution has the same form as in the nonrelativistic
case and the shapes of relativistic and nonrelativistic solitons coincide.

\subsubsection{More stationary solutions}
Encouraged by the results from the nonrelativistic model, one  may try to find stationary solutions with the same charge density as the BPS solutions. In this case however, one would have to use noncommutative Lorenz boosts \cite{BakLee} instead of a Galilean boosts to make the solitons move. We have not worked out the details of this but we feel confident that the solutions can be found.

\subsection{The two soliton case}

\subsubsection{The solution}
As in the nonrelativistic case the two soliton solution can be
written in the form \cite{Bak:2001sg}
\be
\label{2sol:rel}
\phi & = & \lambda\left(\sin\alpha\op{0}{+} + \cos\alpha
\op{1}{-}\right), \\
K & = & z \left(C\, \ket{0}\bra{1} + \frac{1}{C}\, \ket{1}\bra{0}\right)
+ S_2 c S_2^\dg,
\nonumber
\ee
and the BPS equations (\ref{BPS:rel1},\ref{BPS:rel2},\ref{BPS:rel3}) give
\be
\label{2sol:sol}
1-|z|^2\left(C^2 - C^{-2}\right)  =  \frac{2\theta\lambda^2}{\kappa^2}
\sin^2\alpha\left(u^2-\lambda^2\sin^2\alpha\right), \nonumber \\
1+|z|^2\left(C^2 - C^{-2}\right) =  \frac{2\theta\lambda^2}{\kappa^2}
\cos^2\alpha\left(u^2-\lambda^2\cos^2\alpha\right), \\
\cot^2\alpha = C^2\tanh|z|^2~. \nonumber
\ee
Unlike the nonrelativistic case, it is not possible to solve
(\ref{2sol:sol}) in a closed form. It is however straightforward to find
approximate solutions for small and large soliton separations. We find
\be
C & = & |z| + \left(\frac{\theta u^2\lambda_0^2}{\kappa^2}
 - \frac12\right)|z|^5 + {\cal O}\left(|z|^9\right),
\\
\cot\alpha & = & |z|^2 + \left(\frac{\theta u^2\lambda_0^2}{\kappa^2}
- \frac13\right)|z|^6 + {\cal O}\left(|z|^{10}\right),
\\
\lambda^2 & =&  \lambda_0^2 - \frac{2\lambda_0^4}{u^2-2\lambda_0^2}|z|^4
+ {\cal O}\left(|z|^8\right), \\
\lambda_0^2 & = & \frac{u^2}{2}\left(1\pm\sqrt{1-\frac{4\kappa^2}{\theta
u^4}}\right),
\ee
for $|z| \ll 1$ and
\be
C & = & 1 - \frac{\theta\lambda_\infty^2(u^2-\lambda_\infty^2)}
{4\kappa^2|z|^2 - \theta\lambda_\infty^2(2u^2-\lambda_\infty^2)}
{\rm e}^{-2|z|^2} + {\cal O}\left({\rm e}^{-4|z|^2}\right), \\
\cot\alpha & = &1 - \frac{4\kappa^2|z|^2|}
{4\kappa^2|z|^2 - \theta\lambda_\infty^2(u^2-\lambda_\infty^2)}
{\rm e}^{-2|z|^2} + {\cal O}\left({\rm e}^{-4|z|^2}\right), \\
\lambda^2 & = & \lambda_\infty^2 +
{\cal O}\left({\rm e}^{-4|z|^2}\right), \\
\lambda_\infty^2 & = & u^2 \left(1\pm\sqrt{1-\frac{2\kappa^2}{\theta
u^4}}\right),
\ee
for $|z| \gg 1$.

Note that --- modulo different dependence of the
coefficients on $z$ --- the form of $K$ in relativistic and
nonrelativistic cases coincide and the leading terms in the $C$
expansions for small and large $|z|$ are the same. Therefore the
analysis of the shape of the two soliton solution performed in section
\ref{the:shape} can be repeated for the relativistic solution virtually without
any changes.

\subsubsection{The dynamics}
To understand the dynamics of slowly moving solitons we use
Manton's method of moduli space approximation
\cite{Manton:1981mp, Samols:ne} (see also \cite{LRvU,HLRvU}).
The effective action, obtained from (\ref{lagrangian:rel})
by inserting the ansatz (\ref{2sol:rel}) with time dependent moduli $z$
has the form
\be
\label{effa:rel}
S_{\rm eff} = 2\pi\theta\int\!dt\;
\left[g_{zz}\dot z^2 + g_{z\bar z}\dot z\dot{\bar z} +
g_{\bar z\bar z}\dot{\bar z}{}^2
+i{\cal A}\left(z\dot{\bar z} - \dot z\bar z\right)\right]
\ee
with
\be
\label{metric:rel}
g_{zz}
& = & \left(\partial_z\lambda\right)^2 +
\lambda^2\left[\left(\partial_z\alpha\right)^2
-\frac{\bar z{}^2}{4}
\left(\tanh^2|z|^2 \sin^2\alpha + \coth^2|z|^2 \cos^2\alpha\right)
\right] ,\nonumber \\
g_{z\bar z}
& = &
2\left|\partial_z\lambda\right|^2 +
2\lambda^2\left[\left|\partial_z\alpha\right|^2
-\frac{|z|^2}{4}
\left(\tanh^2|z|^2 \sin^2\alpha + \coth^2|z|^2 \cos^2\alpha\right)
\right] \nonumber \\
&& + \; \lambda^2\left[|z|^2 + \tanh|z|^2\sin^2\alpha +
\coth|z|^2\cos^2\alpha\right],  \\
g_{\bar z\bar z} & = & \overline{g_{zz}}~,  \nonumber \\
{\cal A} & = & \frac{\kappa}{\theta \sinh2|z|^2}
\left[\cosh2|z|^2 + |z|^2 \left(C^2 - \frac{1}{C^2}\right)
-\frac12\sinh2|z|^2\left(C^2 + \frac{1}{C^2}\right)\right].
\nonumber
\ee
Using (\ref{2sol:sol}) it is straightforward to express
$\partial_z\lambda$ and $\partial_z\alpha$ in terms of $\lambda, \alpha$ and $C$, but the result is complicated and not particularly illuminating and
we refrain from giving it here. We also not analyze the dynamics
that follows from (\ref{effa:rel}) in detail and discuss only the
limiting cases of small and large $|z|$.

For $|z|\ll 1$ the effective action takes the form
\be
\label{rel:small}
S_{\rm eff} = 2\pi\theta\lambda_0^2\int\!dt\;
r^2\left[3\left(2\dot r^2 + r^2\dot\varphi^2\right)
+ \frac{2u^2}{\kappa}\
r^2\dot\varphi
\right]
+{\cal O}\left(r^6\dot r^2, r^8\right)~,
\ee
where $z = r\,{\rm e}^{i\varphi}$ and we have dropped a total derivative
term. In the coordinates
\[
R = r^2~, \hskip 5mm \vartheta = \sqrt{2}\varphi~,
\]
this action reads
\begin{equation}
\label{rel:small2}
S_{\rm eff} = 3\pi\theta\lambda_0^2\int\!dt\;
  \left(\dot R^2 + R^2 \dot\vartheta^2 + \frac{2\sqrt{2}u^2}{3\kappa}\
R^2\dot\vartheta\right)
\end{equation}
and leads to the equation of motion
\be
\label{eom:small}
\ddot R - R\left(\dot\vartheta + \frac{2\sqrt{2}u^2}{3\kappa}\right)
\dot\vartheta & = & 0~,\\
\frac{d}{dt}\left[R^2\left(\dot\vartheta +
\frac{\sqrt{2}u^2}{3\kappa}\right)\right] & = & 0
\hskip 5mm
\Rightarrow
\hskip 5mm
\dot\vartheta = \frac{c_1}{R^2} - \frac{\sqrt{2}u^2}{3\kappa}~,
\nonumber
\ee
where $c_1$ is an integration constant.

The simplest solution of (\ref{eom:small}) correspond to solitons
circulating around each other with fixed angular velocity,
\[
R = {\rm const}~,
\hskip 5mm
 \dot\vartheta =- \frac{2\sqrt{2}u^2}{3\kappa}~,
\hskip 1cm
\Leftrightarrow
\hskip 1cm
r = {\rm const}~,
\hskip 5mm
\dot\varphi =- \frac{2u^2}{3\kappa}~.
\]
Another simple solution is obtained for $c_1 = 0$,
\[
r^4 = c_2^2\sin^2\left[\frac{\sqrt{2}u^2}{3\kappa}(t-t_0)\right],
\hskip 5mm
\dot\varphi =- \frac{u^2}{3\kappa}~,
\]
and describes head-on scattering of the solitons.

In the general case of $c_1\neq 0$ it
is possible (neglecting a term $\frac{2u^4}{9\kappa^2}r^4$ in
comparison with $c_1^2 r^{-4},$ which is consistent with the
approximation made in deriving (\ref{rel:small})) to write
down the solution in the form
\[
r^4 = c_3^2(t-t_0)^2 + \frac{c_1^2}{c_3^2}~,
\hskip 5mm
\dot\varphi =- \frac{u^2}{3\kappa}+ \frac{1}{\sqrt 2}\
\frac{c_1c_3^2}{c_1^2 + c_3^4(t-t_0)^2}~.
\]

In the region $|z|\gg 1$ one gets
\be
\label{eff:large}
S_{\rm eff} & = & 2\pi\theta\lambda_\infty^2\int\!dt\;
r^2\left(\dot r^2 + 2r^2\dot\varphi^2\right)
+{\cal O}\left({\rm e}^{-2|z|^2}\right).
\ee
In the variables
\[
R = r^2~, \hskip .5cm \vartheta = 2\sqrt{2}\varphi~,
\]
(\ref{eff:large}) takes a form of a free particle action. Consequently
the general solution of the equations of motion that follow from
(\ref{eff:large}) can be presented in the form
\[
r^2{\rm e}^{i2\sqrt{2}\varphi} =
\zeta_0 t + \zeta_1
\]
with (complex) constants $\zeta_0$ and $\zeta_1$.

We thus see that the naive moduli space has conical singularities in
both the short and long distance limit. This is so even for identical
particles where we have to quotient the moduli space by a $Z_2$ action, which for the original GMS solitons removed the apparent conical singularity at
the center of the moduli space \cite{HLRvU,GMSII}. However, the result agrees
with what we found in the non-relativistic case where the phase space
is singular both in the long and short distance limit.

\section{Conclusions}
We have studied one and two soliton solutions in noncommutative
Chern-Simons theory coupled to nonrelativistic and relativistic scalar
fields. We studied the shape of the solitons using the exact Seiberg-Witten
map. We have found new stationary solutions with time-independent charge and energy densities. The most general state could be written as a certain superposition of basis states. We have also integrated the infinitesimal Galilean transformations of \cite{galil}, and used them to generate moving solitons.

For the two soliton solutions we tried to study the dynamics using the moduli space approximation. The direct approach is possible only in the relativistic case because in the nonrelativistic case the action is linear in time derivatives. In the relativistic case, we found explicit expressions for the metric in the limits where the solitons are far apart and when they are close. In these regimes the
metric is flat with a conical singularity at the center. However,
the conical singularity is different in the two limits. This makes the
interactions ``long range'' and the solitons affect each other at
large distances. This awkward long range interaction could be avoided,
but only at the price of an equally awkward redefinition of the
physical variables. In the nonrelativistic case, we could not find a metric, but found a symplectic form that has singularities that resemble the singularities of the metric in the relativistic case. 

Noncommutative theories are known to link UV and IR behavior;
the long range interaction that we find could be related to this, and
could be probed by studying the theory on a cylinder or torus. The
physical meaning of the model should be elucidated by coupling to a
physical external electromagnetic field.  

The new types of solutions require further study. We should investigate the meaning of the gauge-invariant observables that distinguish the basis states with quantized energy from the general mixed states. We should also
determine if such solutions arise in other theories, {\it e.g.}, 
the noncommutative Maxwell-Higgs system.

\section*{Note added:}
An earlier version of this paper found only a special subfamily of the solutions that we discovered here, and were led (falsely) to conclude that no Galilean transformations are possible, and that moving solitons travel with quantized velocities. We thank P.~A.~Horvathy, L.~Martina and P.~C.~Stichel for directing our attention to \cite{galil} and enlightening email exchange on the subject. We have been informed by the authors that they have also written a new paper on related topics \cite{HMS}. We also thank D.~Bak for useful communications.

\section*{Acknowledgments:}
We are grateful to Hans Hansson and Anders Karlhede for illuminating
discussions. We thank our respective institutions for generously
hosting the other members of the collaboration at various times during
this project.  LH and UL are grateful for the stimulating atmosphere
at the Simons Workshop in Mathematics and Physics at the C.N. Yang
Institute for Theoretical Physics, Stony Brook, where this work was
completed. The work of LH was supported by the EC IHP network
HPRN-CT-1999-000161. UL acknowledges support in part by EU contract
HPNR-CT-2000-0122 and by VR grant 650-1998368. The work of MR was
supported in part by NSF Grant No. PHY-0098527 and Supplement for
International Cooperation with Central and Eastern Euorpe PHY 0300634. 
The work of RvU was supported by the Czech Ministry of Education 
under Contracts No. 143100006, GA\v{C}R 201/03/0512, and ME649.

\end{document}